\documentclass[twocolumn,aps]{revtex4}
\usepackage{epsfig}
\usepackage{amsmath}

\newcommand{\be}{\begin{equation}}
\newcommand{\ee}{\end{equation}}
\newcommand{\bea}{\begin{eqnarray}}
\newcommand{\eea}{\end{eqnarray}}

\def\two{2}
\def\three{3}

\begin{document}
\title{Occam's razor meets WMAP}
\author{Jo\~ao Magueijo$^{1,2,3}$ and Rafael D. Sorkin$^{1,4}$}

\affiliation{$^1$Perimeter Institute for Theoretical Physics, 31
Caroline St, Waterloo N2L 2Y5, Canada\\
$^2$ Canadian Institute for Theoretical Astrophysics,
60 St George St, Toronto M5S 3H8, Canada\\
$^3$ Theoretical Physics, Imperial College, Prince Consort
Rd., London SW7 2BZ, England\\
$^4$ Department of Physics, Syracuse University, Syracuse NY
13244-1130, USA}

\begin{abstract}
Using a variety of quantitative implementations of Occam's razor
we examine the low quadrupole, the ``axis of evil'' effect and
other detections recently made appealing to the excellent WMAP
data. We find that some razors {\it fully} demolish the much
lauded claims for departures from scale-invariance. They all
reduce to pathetic levels the evidence for a low quadrupole (or
any other low $\ell$ cut-off), both in the first and third year
WMAP releases. The ``axis of evil'' effect is the only anomaly
examined here that survives the humiliations of Occam's razor, and
even then in the category of ``strong'' rather than ``decisive''
evidence. Statistical considerations aside, differences between
the various renditions of the datasets remain worrying.
\end{abstract}

\pacs{PACS Numbers: *** }
\date{\today}

\maketitle

\section{Introduction}
A better fit to the data can always be obtained by appealing to a
theory containing more free parameters. The extra knobs can't
harm, and quite often help the job of fitting data. Intellectual
honesty, however, tells us that a better fit may then not signal
evidence for the theory, but merely unfair advantage over its
competitors. Confronted with two theories fitting the data equally
well we'd prefer the simpler one, the theory containing
fewer parameters or based on a less complicated model.

Such considerations form the basis of Occam's razor, but a quantitative
formulation is notoriously hard to come by.  It's clear that the real
``evidence'' should combine the naive goodness of fit with a penalty
function measuring the complexity of the theory.  But several distinct
rationales for doing this may be found in the literature, notably the
Akaike~\cite{aic} and Bayesian~\cite{bic} information criteria (AIC and
BIC) and the Turing machine based criterion proposed by one of
us~\cite{raf}.  Simplicity, it seems, is in the eye of the beholder.

Furthermore, subjective double standards seep into the analysis,
and the rigors of penalization are often reserved to results one
doesn't like. For example, the CMB community has resisted applying
Occam's razor to inflationary parameters (see~\cite{lid,ba1,ba2}
for notable exceptions) and to some power spectrum
features~\cite{topo,lowcls}; but with reference to anomalies
unpalatable to just about everyone (such as the ``axis of evil''
effect, the embarrassing statistical anisotropy exhibited on the
largest angular scales~\cite{ev1,ev2,ev3}), the strictest
penalization is enforced~\cite{wmap3c}.
(The criterion employed therein to scrutinize the axis of
evil effect is loosely the AIC.)

We applaud this type of application of Occam's razor, but we believe it
should be employed impartially.    The purpose of the present paper is
to examine
some of the proposed
Occam razors, and to apply them democratically to both ``likable'' and
``undesirable'' features in the large-angle CMB anisotropy.  We
examine the WMAP data~\cite{wmap1,wmap3a,wmap3c}, in its first and
third year releases, and in various renditions dealing differently
with the galactic plane.
We focus on
claims for departures from
scale invariance and for reionization (Section~\ref{brands}),
the evidence for a low quadrupole and a low $\ell$ power cut-off
(Section~\ref{underpower}),
and the strength of the detection of
the so called axis of evil effect (Section~\ref{evil}).

\section{Brands of Occam's razor}\label{brands}
We first review some well-known criteria for evidence,
adopting a notation similar to that of \cite{raf}.
Let ${\cal L}$
be the likelihood and $k$ the number of parameters of the model.
They will be tuned so as to maximize the likelihood or,
equivalently, minimize the information $I$.  The information in the
data given the theory is defined as minus the logarithm of the
likelihood.
But in fact we want to minimize the information in the
data {\it and} the theory together,
that is: %
\be I(D,T)=I(D|T)+I(T)\ee%
so that $I(T)$ is the
penalty
referred to above.

According to some authorities,
strong evidence for a theory over a ``base model'' requires
an improvement in $I(D,T)$ by at least 3
(see~\cite{jef,muk}).  The title of ``decisive evidence'' is not
normally bestowed unless the improvement exceeds 5.

All the razors we will wield fit into the above scheme,
but they differ
in how they define $I(T)$.
According to the Akaike information
criterion (AIC)
the information in
a theory is simply its {\it number} of parameters, so that %
\be I_A(D,T)= -\ln {\cal L} + k.\ee%
This is obtained by an approximate minimization of the
Kullback-Liebler information entropy.

Rather different is the
the Bayesian information criterion (BIC),
based on the penalty%
\be I_B(T)={k\over 2}\ln N\ee%
where $N$ is the number of data points
being fit.
It results from an
approximation to the true Bayesian evidence, giving the model a
uniform prior.  The full Bayesian evidence, where one integrates
the likelihood over the full set of parameters, has also been
considered.

The criterion developed in~\cite{raf} interprets $I(T)$ entropically and
algorithmically.
It estimates the information in a theory $T$ in terms of the
number of bits of ``memory'' required to {\it store} the
parameters.  From this point of view, a theory's complexity depends not
only on how many parameters it contains, but also on the precision with
which they are stored.  The resulting penalty term $I(T)$ is not simply
a function of $N$ and $k$ but depends on the details of the theory.
Typically it includes a term equal to (\three) as well as other terms that
ensure that a theory with more parameters than data points will never be
judged a good fit.  The advantage this approach has over AIC and BIC is
that it never has to appeal to the asymptotic approximation $N\gg1$.
It's disadvantage
is that it is harder to apply since $I(T)$ is not a simple universal
function of $N$ and $k$.
In the spirit of the abbreviations ``AIC'' and ``BIC'' we will refer to
this third razor as ``HIC'' or simply ``$H$'',
since the differential goodness of fit is denoted by $H$ in \cite{raf}.

The various criteria do not always agree, even qualitatively.
Take the statements that WMAP displays strong evidence for
reionization and a spectrum of density fluctuation that is not
scale invariant~\cite{wmap3c}.
Both assertions rely on an improvement to the fit
$\Delta I(D|T)\approx \Delta\chi^2/2= -4$
(see Table 3 in~\cite{wmap3c})
and for both, this costs an extra parameter.
Using AIC we get  $\Delta I(D,T)=-3$ pointing
toward a detection. But $N$ is between 1500 and 3100, so $\ln N$
is around 8. Using BIC this implies $\Delta
I(D,T)\approx 0$, most definitely not a detection.
We have not worked out the HIC value of $I$, but we would expect it to
resemble BIC more than AIC, leading again to the verdict of ``no
detection''.
Of course
one cannot drop both departures from scale invariance and
reionization and there are strong astrophysical reasons for
preferring reionization to a tilted spectrum.  Therefore based on
WMAP it seems prudent to say that there is no strong evidence for
$n_S\neq 1$.

\section{Is the quadrupole underpowered?}\label{underpower}
Much attention has been paid to the low power observed in the
lowest multipoles ($\ell=2$ in particular), but how strong is the
evidence when shaved with Occam's razor?  This is essentially a
problem of variance estimation.  Given a sample and an externally
inferred variance $\sigma^2_E$, when is it worth revising
$\sigma^2_E$ in the light of the sample?  Here $\sigma_E^2$  is
obtained by appealing to a theory of the whole spectrum, dependent
only on a small number of parameters (e.g. $\Omega$ and
$\Omega_\Lambda$). These are fixed primarily by the higher
multipoles (the Doppler peaks), so as far as the low multipoles
are concerned $\sigma^2_E$ is external.

The ``null hypothesis'' H0 is that $\sigma^2_E$ is correct, and the
observed low power a fluke.  Since the costs of estimating
$\sigma^2_E$ are borne elsewhere, $I(T)=0$ and
$I(D,T)=I(D|T)$.  The catch is that the fit to the data is far from
perfect.
Introducing the ``observed variance'' of the sample,%
\be \sigma^2_S={1\over N}\sum x^2_i \ee%
we have%
\be I(D,T)=-\ln P(D|T)={N\over 2}
 {\left[\ln\sigma^2_E +{\sigma^2_S\over \sigma^2_E}\right]} \ee %
far from its minimum.

The alternative hypothesis H1 is that the power is indeed low
and that $\sigma^2_E$ should be replaced by an internal estimate,
$\sigma^2_I$, obtained using the sample and bearing its costs.
The procedure for applying HIC can be adapted from~\cite{raf}
and goes as follows (the only novelty is that
here the average is known).
Firstly, we minimize $I(D|T)$, with solution
$\sigma^2_I=\sigma^2_S$. This cannot be stored to infinite
accuracy, so we expand around the minimum:
\be\label{form1}
  I(D|\sigma_S,\Delta\sigma)={N\over 2}{\left[\ln\sigma^2_S
  +1\right]} +N{\left(\Delta\sigma\over\sigma_S\right)}^2.
\ee
Averaging over a uniform distribution in
$\Delta\sigma\in(-\delta\sigma/2,\delta\sigma/2)$ gives  $\langle
\Delta\sigma^2\rangle=\delta\sigma^2/12$, so that:%
\be
  \label{form11} I(D|\sigma_S,\delta\sigma)={N\over
  2}{\left[\ln\sigma^2_s +1\right]} +{N\over
  12}{\left(\delta\sigma\over\sigma_S\right)}^2 .
\ee
The
storage penalty, on the other hand, is  %
\be I(T)=-\ln{\delta\sigma\over \sigma_S}\ee %
so $I(D,T)$ is minimized for optimal accuracy: %
\be
\delta\sigma={\sqrt{6\over N}}\sigma_S .\ee%
Thus the information in the data and H1 is %
 \be
I(D,T)={N\over 2}[\ln\sigma_S^2 +1] +{1\over 2} -\ln
\sqrt{6\over N} \ee %
The evidence $H$ against the null hypothesis H0 is the difference
between its information and that in H1 (positive $H$ favors H1).
This may be written as $H=H_f - H_p$, where
$H_f$ is the improvement in
the fit%
\be
H_f={N\over 2}{\left[\ln{\sigma_E^2\over
\sigma_S^2}+{\sigma_S^2\over \sigma_E^2} -1\right]}
\ee %
(this is often approximated by  $-\Delta\chi^2/2$),
and $H_p$,
the penalty paid by H1 for introducing a new parameter,
is
\be \label{pen1}%
  H_p={1\over 2} \left[1 + \ln {N\over 6}\right] \approx
{1\over 2}\ln N - 0.4
\ee
An exact rendition of this argument (not appealing to
Taylor expansion (\ref{form1})) leads to penalty
\be \label{pen2}%
  H_p={1\over 2} \left[\ln {N-1\over 6} +N\ln{N\over N-1 }  \right] + \psi(N)
\ee
where $\psi(N)$  is a small negative correction, monotonic in $N$,
that never exceeds 0.2 in magnitude and is totally negligible for $N>5$
(for example $\psi(10)=-0.03$).
The AIC would instead quote
$H^{AIC}_p=1$ (with $H^{AIC}=H_f-H_p^{AIC}$),
whereas the BIC would introduce:
\be%
 H^{BIC}_p={1\over 2}\ln N  \ee %
(with $H^{BIC}=H_f-H_p^{BIC}$) which in the large $N$ limit is the
same as (\ref{pen1}) (or (\ref{pen2})) plus constant 0.4.
Generalization for many independent parameters is straightforward.

\begin{table}
\caption{Evidence for a low quadrupole, based on various datasets and
Occam's razors $H$, AIC and BIC.}
\label{quadtab}
\begin{tabular}{|l|l|lll|}
\hline
Map& $H_f$ & $H$ &$H^{AIC}$ & $H^{BIC}$ \\
\hline
ILC1 & 2.47 &2.11 &1.47 &1.67\\
TOH& 2.62 & 2.26 & 1.62 & 1.81\\
DILC & 2.08 & 1.72 & 1.08 & 1.27 \\
WMAP3& 2.32 & 1.96 & 1.32 & 1.51 \\
\hline
\end{tabular}
\end{table}
In Table~\ref{quadtab} we examine the evidence for a low quadrupole.
We consider the first year data as in~\cite{wmap1} (ILC1) and
in~\cite{toh} (TOH), as well
as the third year release~\cite{wmap3a}, both the
debiased internal linear combination map
(DILC) and the MLE estimate (WMAP3). Clearly under Occam's razor
we can never claim a significant detection, whatever the dataset.
Adding the octupole and other low $\ell$ does little to improve the
situation. Visual inspection of the plots
in~\cite{wmap3a} shows that many of these low $\ell$
``anomalies'' have disappeared in the three year data.
But they were never significant, as the analysis of the first year data
presented in Table~\ref{lowcltab} shows. Naturally $H_f$ improves
as more and more multipoles are considered, but these bring in
new parameters and so the associated ``detections'' are erased
under the weight of Occam's razor.
This table refers to first year TOH; in  other datasets/renditions the
evidence is even lower. By bringing more $\ell$s into the analysis
the evidence decreases further.

None of this will surprise several
authors~\cite{efst,toh,gazt,slosar,biel}; yet, to drive the point
home we stress that the evidence for a low quadrupole  -- bad as
it is -- is still stronger than the evidence for a non scale
invariant spectrum under the BIC. Also the message has yet to
fully filter to enthusiastic theorists (e.g.~\cite{topo}). For
example claims have been made~\cite{carroll} that DGP
gravity~\cite{dgp} fits better the low $\ell$ spectrum. While it
might be true that the theory achieves a better fit without
introducing new parameters (and therefore doesn't fall prey of
{\it further} penalties) the fact remains that it corrects a
misfit that is not significant to begin with.

\begin{table}
\caption{Evidence (or lack thereof) for low power at small $\ell$ using
the most sympathetic dataset (TOH).}
\label{lowcltab}
\begin{tabular}{|l|l|lll|}
\hline
$\ell$ or $\ell$ range& $H_f$ & $H$ &$H^{AIC}$ & $H^{BIC}$ \\
\hline
2 & 2.62 & 2.26 & 1.62 & 1.81\\
3&0.35&-0.18&-0.65&-0.62\\
2-3& 2.98& 2.08 & 0.97 &1.19\\
4&1.18&0.51&0.18&0.09\\
2-4& 4.16& 2.59 &1.15  &1.28 \\
\hline
\end{tabular}
\end{table}

\section{The axis of evil}\label{evil}
Many paths lead to the axis of evil. Planarity
statistics~\cite{ev1}, Maxwell multipole vectors~\cite{ev2,max},
and $m$-preference statistics~\cite{ev3} are examples. Here we
focus on the planarity of $\ell=2,3$, that is, the fact that in
the frame pointing to $(b,l)\approx (60,-100)$ in Galactic
coordinates, the power is concentrated in the $m=\pm\ell$ modes.
How seriously should we take this?

The more abstract estimation problem is: when is it justified
splitting the $a_{\ell m}$ sample into sub-samples with different
variances? This is a variation on the calculation in the previous
section with a subtlety: the result is frame dependent. Consider a
sample with $N$ elements and sample variance $\sigma^2_S$ (the
$2\ell+1$ modes of a multipole), and two sub-samples with $N_1$
and $N_2$ elements and sample variance $\sigma^2_{S1}$ and
$\sigma^2_{S2}$ (the planar modes $m=\pm\ell$, and all the
others). The difference in $I(D|T)$ between the null hypothesis
(don't split the sample) and the alternative hypothesis (split) is
\be%
H_f=\ln {\sigma_{S}^N\over \sigma_{S1}^{N_1}\sigma_{S2}^{N_2}}\ee%
where $N=N_1+N_2$ and
$N\sigma_S^2=N_1\sigma_{S1}^2+N_2\sigma_{S2}^2$.
This depends only on the suppression ratio \be
\epsilon={\sigma^2_{S2}\over
\sigma^2_{S1}}\ee %
and therefore one can consider the issue of planarity even if the
evidence for an internal $\sigma^2_S$ is small or nonexistent. The
value of $\epsilon$ depends on the $z$-axis coordinates $(b,l)$,
which should be chosen to maximize $H_f$. In the process we add
two more parameters to the Occam's razor bill.

\begin{table}
\caption{The planarity of the $\ell=2,3$ modes using TOH (top
rows) and WMAP3 (bottom). There is nothing anomalous with the
planarity of  $\ell=2$ and $\ell=3$, taken on their own. It's the
fact that the planarity occurs in roughly the same direction (and
with roughly the same suppression ratio $\epsilon$) for both
multipoles that substantiates the anomaly.} \label{eviltab}
\begin{tabular}{|l|l|lll|l|ll|}
\hline
Data&$\ell$s& ($b$ & $\quad l$) & $\epsilon$  &$H_f$  &$H^{AIC}$ & $H^{BIC}$ \\
\hline
&2 &58 & -103 &.030 & 3.09   &0.09 & 0.68  \\
TOH &3& 62 & -121 &.025 & 5.06   & 2.06 & 2.14\\
& 2-3 & 61& -113 & .032 & 7.48  & 4.48 & 3.76 \\
\hline
&2 &70&-127&.036 &2.84 &   -0.16 & 0.43 \\
WMAP3&3&62 &-122 & .035 & 4.29    & 1.29 &1.37  \\
&2-3  &64 &-123 & .038 & 6.89   & 3.89  & 3.16 \\
\hline
\end{tabular}
\end{table}
This is the procedure adopted for analyzing  each multipole
independently and in Table~\ref{eviltab} we present results for
two datasets: TOH and the WMAP three year data. We find that $H_f$
is around 3 for $\ell=2$ and 5 for $\ell=3$, at the cost of
introducing 3 parameters (the axis and the ratio of power
$\epsilon$) for each multipole. Using AIC this degrades $H_f$ to a
$H$ around 0 and 2, respectively. Results for the BIC are reported
in the same table. As in previous studies~\cite{ev1,ev3} we find
no serious evidence for an anomaly if {\it each multipole is taken
on its own}. Given a random, statistically isotropic multipole
there is always a frame in which most of the power is concentrated
in a single $m$; that this $m$ equals $\ell$ is not unlikely for
small $\ell$.

What turns the axis of evil into a menace is that the maximal
$H_f$ for $\ell=2$ and $\ell=3$
is reached with roughly the  same parameters (see values in Table
III). Thus if we take a single axis and $\epsilon$ chosen so as to
maximize the total $H_{Tf}=H_{Qf}+H_{Of}$, we obtain a $H_{Tf}$
only slightly worse than the sum of the separate optimal $H_{Qf}$
and $H_{Of}$; the parameter cost, however, is halved. Our results
are described in Table III. The search for the joint axis was done
numerically, and we see that the result is heavily weighed by the
octupole. The common $\epsilon$ was found via the method of
Lagrange multipliers, i.e.
by maximizing%
\be H_{Tf}=H_{Qf}+H_{Of}-\lambda
[\sigma^2_{Q1}\sigma^2_{O2}-\sigma^2_{O1}\sigma^2_{Q2}]\ee%
with solution:%
\bea%
\sigma^2_{Qi}&=&{\sigma_{SQi}^2\over 1 \pm A/2}\nonumber\\
\sigma^2_{Oi}&=&{\sigma_{SOi}^2\over 1 \mp A/2}\nonumber
\eea%
where $i=1,2$ indexes the sub-samples and $A$ is the solution of a
quadratic equation expressing $\epsilon_O=\epsilon_Q$ (an equation
that only depends on the sample ratio
$\epsilon_{SO}/\epsilon_{SQ}$.)

As shown in Table III our evidence for an anomaly is always above
$H=3$, i.e. ``strong evidence''.  One may therefore wonder where
is the discrepancy with the analysis in~\cite{wmap3c}? In that
work the axis of evil was modeled as a modulation by an
underlying large-scale function, and a model was found with
$H_f=4$ (a chi-squared improvement of 8) at a cost of 8
parameters. Using either AIC or BIC the value of $H$ is therefore
negligible. However, here we exhibited a model improving the fit
by about $H_f=7$ at a cost of 3 parameters. This
(phenomenological)
model is simply based on a diagonal covariant matrix for $\ell=2,3$
of the form:%
\be\label{model}%
{\langle |a_{\ell m}|^2 \rangle}({\mathbf n}) =c_\ell
(\delta_{\ell |m|}+ \epsilon (1-\delta_{\ell |m|}))
\ee%
Hence the poor evidence reported in~\cite{wmap3c} is not a
deficiency of the axis of evil effect or the data, but merely a
shortcoming of the proposed model itself.  One can always find a
model for any anomaly containing a number of parameters so
large as to drive $H$ down to a small value.
But the issue is: what is the
value of $H$ for the best model of that anomaly, the model
with the optimal trade off between fit and number of parameters?
We have gone a fair way
toward answering this question.

\section{Conclusions}
In this paper we subjected to some of Occam's razors three patterns that
people have claimed to see in the CMB data:
departures from scale invariance,
a low quadrupole, and
the anisotropy that has come to be known as the ``axis of evil''.
Specifically, we considered the razors that we called AIC, BIC and
HIC.  All three agreed to discount the claim for a low quadrupole, while
in contrast, the two that we brought to bear on the axis of evil both
suggested that it should be taken seriously.  Only in relation to
scale-invariance was there disagreement, with AIC tending to accept the
claim and BIC definitely rejecting it.  (We did not consult HIC in
connection with the first and third effects, but we plan to do so in a
later version of this preprint.)

It is somewhat embarrassing that Occam razors
can disagree,  but a glance at equations
(\two) and (\three) reveals that
this is inevitable, since the penalty terms $N$ and $\ln\sqrt{N}$ are
very different when the number of data points $N$ is $\gg1$.
By comparing these two expressions, one
sees that BIC will be more lenient
than AIC when $N$ is small, but much tougher when $N$ is big (the
crossover coming around $N=7$).
For HIC, it is harder to make a blanket statement, but experience has
shown that it tends to agree more closely with BIC, probably since each
relies, in its own way, on a version of Bayes' rule.

In the case of the claimed departure from scale-invariance, we would
thus expect HIC to agree with BIC in favoring a negative verdict, which
at the very least should be added as a word of caution to the
conclusions reported in~\cite{wmap3c}.
By way of comparison, it's worth pointing out that, even if we accept
the more favorable value of $H$ coming from AIC, the evidence for scale
non-invariance is no better than that for the ``axis of evil''.
When all razors agree on
a lack of evidence, as is the case with the underpowered
quadrupole, one should definitely not lose sleep over the
anomaly, and we hope keen theorists will divert their creativity
elsewhere.

But
even when different razors agree on an anomaly -- such as
the axis of evil -- one should not trust the result blindly.
The
issue of systematics remains of paramount importance, as shown by
the significant differences in $H$ obtained from the various
datasets and methodologies used to deal with the galactic
foregrounds.
And one should bear in mind that even the most enthusiastic
``Ockhamist'' would be
unlikely to claim for his or her favorite razor a freedom
from ambiguity\footnote
{The most important ambiguities are those that affect the value of
 $I(T)$.  Do we treat the algorithm as storing $\sigma$ or $\sigma^2$,
 for example?}
better than  $\Delta{H}=\pm0.3$ or so.
In addition it's probably fair to say that the
trouble of rewriting cosmology textbooks deserves in itself a
penalty factor.  This is hard to evaluate but it may translate
into the requirement of a higher level of evidence than
``strong'', at the phenomenological level.  Perhaps the ever
improving polarization maps will have a say on the matter and tilt
the scales.  This issue is currently being very actively
investigated.



{\bf Acknowledgments} We'd like to thank Carlo Contaldi, Olivier
Dor\'e and Kate Land for discussion and comments. This research
was partly supported by NSF grant PHY-0404646.

\label{lastpage}


\begin{thebibliography}{99}
\bibitem{aic}H. Akaike, IEEE Trans. Auto. Cont., 19, 716, 1974.
\bibitem{bic}G. Schwarz, Annals of Statistics, 5, 461, 1978.
\bibitem{raf}R. Sorkin, Int. J. Theor. Phys. 22, 1091, 1983.
\bibitem{lid}A. Liddle, Mon. Not. Roy. Astron. Soc., 351, L49-L53, 2004.
\bibitem{ba1}M. Hobson, S. Bridle and O. Lahav, MNRAS 335, 377, 2002.
\bibitem{ba2} A. Niarchou, A. Jaffe and L. Pogosian, Phys. Rev. D 69,
063515, 2003.
\bibitem{jef}H. Jeffreys, Theory of probability, Oxford University Press, 1961
(Appendix B).
\bibitem{muk}S. Mukherjee et al, ApJ 508, 314 (1998).
\bibitem{topo}J. Luminet et al,  Nature 425: 593, 2003
\bibitem{lowcls}C. Contaldi et al, JCAP 0307: 002, 2003.
\bibitem{ev1} A. de Oliveira-Costa et al,  Phys. Rev. D69 (2004) 063516.
\bibitem{ev2}D. Schwarz et al, Phys.Rev.Lett. 93 (2004) 221301.
\bibitem{ev3} K. Land, J. Magueijo, Phys. Rev. Lett. 95: 071301, 2005;
 astro-ph/0502237.
\bibitem{ev4}K. Land, J. Magueijo,
Phys.Rev. D72 (2005) 101302; astro-ph/0507289.
\bibitem{wmap3c}D. Spergel et al, astro-ph/0603449.
\bibitem{wmap3a}G. Hinshaw et al, astro-ph/0603451.
\bibitem{wmap1}Bennett C.L. et al., 2003, Astrophys. J. Suppl, 148, 1.
\bibitem{toh}M. Tegmark, A. de Oliveira-Costa, A. Hamilton,
Phys.Rev. D68 (2003) 123523.
\bibitem{efst}G. Efstathiou et al, Mon.Not.Roy.Astron.Soc. 346 (2003) L26.
\bibitem{slosar} A. Slosar and U. Seljak,  Phys.Rev. D70 (2004).
\bibitem{biel}P. Bielewicz et al, Astrophys. J. 635: 750-760,
2005.
\bibitem{gazt}E. Gaztanaga et al, Mon.Not.Roy.Astron.Soc. 346 (2003) 47-57.
\bibitem{dgp}G. Dvali, G. Gabadadze, M. Porrati,
Phys.Lett.B485: 208-214, 2000.
\bibitem{carroll} I. Sawicki and S. Carroll, astro-ph/0510364.
\bibitem{max}K. Land and J. Magueijo, MNRAS 362, L16-L19, 2005;
MNRAS, 362: 838-846, 2005.
\end{thebibliography}
\end{document}